\RequirePackage{fixltx2e}
\documentclass[journal=jpcafh,manuscript=review,layout=onecolumn]{achemso}
\usepackage{graphicx} 
\usepackage{color}
\usepackage{dcolumn}
\usepackage{amsmath,amsfonts} %
\usepackage{braket} %
\usepackage[acronym]{glossaries}
\usepackage[breaklinks, %
colorlinks, %
linkcolor=blue, %
citecolor=blue, %
urlcolor=blue]{hyperref}

\def\be{\begin{equation}}
\def\ee{\end{equation}}
\def\bea{\begin{eqnarray}}
\def\eea{\end{eqnarray}}
\newcommand{\fig}[1]{Fig.~\ref{#1}}

\newcommand{\eq}[1]{Eq.~\eqref{#1}}
\newcommand{\Eq}[1]{Equation~\eqref{#1}}
\newcommand{\eqs}[1]{Eqs.~\eqref{#1}}
\newcommand{\mat}[1]{\ensuremath{\boldsymbol{#1}}}
\newcommand{\vect}[1]{\ensuremath{\vec{#1} }}
\newcommand{\op}[1]{\ensuremath{{#1}}}
\renewcommand{\ket}[1]{\ensuremath{\left|#1\right\rangle}}
\renewcommand{\bra}[1]{\ensuremath{\left\langle #1\right|}}
\renewcommand{\braket}[2]{\ensuremath{\left\langle #1|#2\right\rangle}}
\newcommand{\braOket}[3]{\ensuremath{\left\langle #1\left|#2\right|#3\right\rangle}}
\renewcommand{\Re}{\operatorname{Re}}
\renewcommand{\Im}{\operatorname{Im}}
\newcommand{\Dim}{\mathcal{D}}
\newcommand{\Liou}{\mathcal{L}}
\newcommand{\K}{\mathcal{K}}

\newcommand{\tr}{\ensuremath{\mathrm{tr}}}

\title{Nonadiabatic Quantum Dynamics with Frozen-Width Gaussians}

\author{Lo{\"i}c Joubert-Doriol} %
\affiliation{Department of Physical and Environmental Sciences, University of Toronto Scarborough, Toronto, Ontario, M1C 1A4, Canada} %
\alsoaffiliation{Chemical Physics Theory Group, Department of Chemistry, University of Toronto, Toronto, Ontario M5S 3H6, Canada} %
\author{Artur F. Izmaylov} %
\email{artur.izmaylov@utoronto.ca}
\affiliation{Department of Physical and Environmental Sciences, University of Toronto Scarborough, Toronto, Ontario, M1C 1A4, Canada} %
\alsoaffiliation{Chemical Physics Theory Group, Department of Chemistry, University of Toronto, Toronto, Ontario M5S 3H6, Canada} 

\newacronym{RMSD}{RMSD}{root mean square deviation} %
\newacronym{BH}{BH}{Born--Huang} %
\newacronym{TDSE}{TDSE}{time-dependent Schr\"odinger equation} %
\newacronym{BO}{BO}{Born--Oppenheimer} %
\newacronym{MQC}{MQC}{mixed quantum-classical} %
\newacronym{DFT}{DFT}{density functional theory} %
\newacronym{GP}{GP}{geometric phase} %
\newacronym{CI}{CI}{conical intersection} %
\newacronym{CS}{CS}{coherent states} %
\newacronym{LVC}{LVC}{linear vibronic coupling}
\newacronym{NAC}{NAC}{nonadiabatic coupling} %
\newacronym{FSSH}{FSSH}{fewest-switches surface hopping} %
\newacronym{QCL}{QCL}{quantum-classical Liouville} %
\newacronym{WT}{WT}{Wigner transform} %
\newacronym{DBOC}{DBOC}{diagonal Born--Oppenheimer correction} %
\newacronym{vMCG}{vMCG}{variational Multiconfiguration Gaussian} %
\newacronym{MCE}{MCE}{multiconfiguration Ehrenfest} %
\newacronym{AIMS}{AIMS}{ab inito multiple spawning} %
\newacronym{MCA}{MCA}{moving crude adiabatic} %
\newacronym{NOSSE}{NOSSE}{non-stochastic open system Schr\"odinger equation} %
\newacronym{QME}{QME}{quantum master equation} %
\newacronym{eMCA}{eMCA}{exact-MCA} %
\newacronym{MCTDH}{MCTDH}{multiconfiguration time-dependent Hartree} %
\newacronym{DGAS}{DGAS}{diabatic Gaussians on adiabatic states} %
\newacronym{TDVP}{TDVP}{time-dependent variational principle} %
\newacronym[longplural={degrees of freedom}, %
firstplural={degrees of freedom (DOF)}, plural={DOF}]{DOF}{DOF}{degree
  of freedom} %
\newacronym[longplural={equations of motion}, %
firstplural={equations of motion (EOM)}, %
plural={EOM}]{EOM}{EOM}{equation of motion} %

\begin{document}

\renewcommand{\abstractname}{Abstract}
\renewcommand{\tocentryname}{}


\begin{tocentry}
\center
\includegraphics[width=5cm]{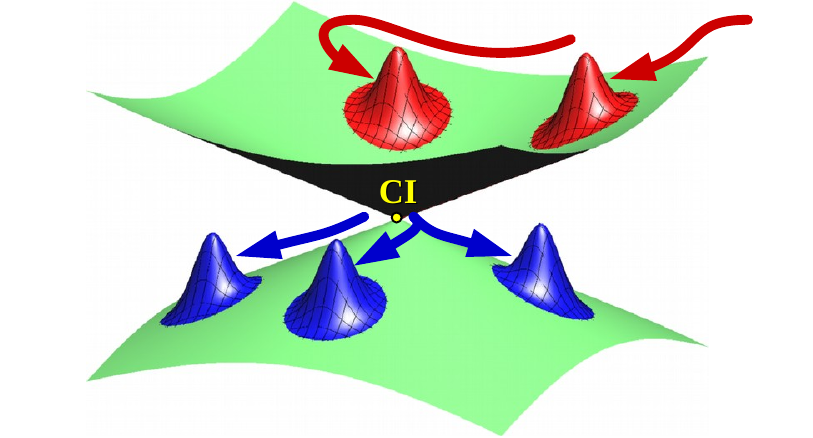}
\end{tocentry}
\sloppy
\begin{abstract}
\noindent We review techniques for simulating fully quantum nonadiabatic dynamics
using the frozen-width moving Gaussian basis functions to represent the nuclear wavefunction.
A choice of these basis functions is primarily motivated by the idea of the on-the-fly dynamics that 
will involve electronic structure calculations done locally in the vicinity of each Gaussian center 
 and thus avoiding the ``curse of dimensionality'' appearing in large systems. For quantum dynamics involving multiple 
electronic states there are several aspects that need to be addressed. First, the choice 
of the electronic state representation is one of most defining in terms of formulation of resulting 
equations of motion. We will discuss pros and cons of the standard adiabatic and 
diabatic representations as well as the relatively new moving crude adiabatic representation. Second,
if the number of electronic states can be fixed throughout the dynamics, the situation is different for the 
number of Gaussians needed for an accurate expansion of the total wavefunction. The latter increases 
its complexity along the course of the dynamics and a protocol extending the number of Gaussians 
is needed. We will consider two common approaches for the extension: 1) spawning and 2) cloning. 
Third, equations of motion for individual Gaussians can be chosen in different ways, 
implications for the energy conservation related to these ways will be discussed. 
Finally, to extend the success of moving basis approaches to quantum dynamics of open systems we 
will consider the Non-stochastic Open System Schr\"odinger Equation (NOSSE).  
\end{abstract}

\glsresetall

\maketitle

\section{Introduction}
\label{sec:introduction}

Quantum dynamics simulations are commonly used to understand microscopical details of ultrafast molecular processes initiated by interaction of the system with UV or visible light.~\cite{Bakulin:2015/nc/16,Rao:2016/jpca/3286,Zheng:2016/jpcc/1375,Wang:2014/prl/113007,Li:2017/nc/453,Vacher:2017/prl/083001,Wu:2015/jcp/074302,Kirrander:2016/jctc/957,Makhov:2016/fd/81,Curchod:2018/cr}
Such ultrafast processes involve two or more electronic states and often the nuclear dynamics takes place  
in areas where potential energy surfaces (PESs) approach each other or cross. These proximities of adiabatic PESs 
lead to break-down of the Born-Oppenheimer approximation and to molecular dynamics beyond a single 
electronic state: nonadiabatic dynamics. This work will consider fully quantum approaches to modeling 
nonadiabatic processes. 

The problem that will be addressed here is solving the \gls{TDSE}, $H\Psi(r,R,t) = i\partial_t \Psi(r,R,t)$,
where $H = T_n+H_e$ is a molecular Hamiltonian that contains the nuclear kinetic energy, $T_n$, 
and the electronic Hamiltonian, $H_e = T_e+V_{ee}+V_{en}+V_{nn}$, consisting of the 
electronic kinetic energy, $T_e$, and all Coulomb potential energies. 
 The electron-nuclear wavefunction, 
$\Psi(r,R,t)$, has a collection of electronic and nuclear variables $r$ and $R$, respectively.   
In order to have nontrivial quantum dynamics revealing properties of the system,
it will be considered that the system is prepared in some non-stationary (or non-equilibrium for open systems) 
state. We will not go into the details of the initial state preparation, mainly assuming some ultra-fast 
excitation available in modern ultra-fast laser spectroscopy. 

The main topic of the current work is approaches that simulate $\Psi(r,R,t)$ in the form 
\bea\label{eq:gwp}
\Psi (r,R,t) & = & \sum_{k=1}^{N_G}\sum_{s=1}^{N_S} C_{ks} (t) g_k(R,t) \phi_s (r),
\eea
where a linear combination of $N_G$ frozen-width Gaussian (FWG) functions $g_k(R,t)$ 
used to describe interstate dynamics within a manifold of $N_S$ electronic states $\phi_s (r)$.
The main motivation behind using FWG is an attempt to defeat the ``curse of dimensionality'' of 
quantum mechanics. Indeed, describing a nuclear part of the wavefunction everywhere would 
require an exponentially large number of basis functions or grid points, while FWGs are local and 
travel only to the most important configurations of the nuclear geometry. 
 We will consider several choices for the electronic states 
$\phi_s(r)$ in this work, but they all can be obtained locally in the vicinity of the corresponding 
FWG centers using well developed techniques of solving the electronic structure problem.  

\begin{figure}
  \centering
  \includegraphics[width=0.5\textwidth]{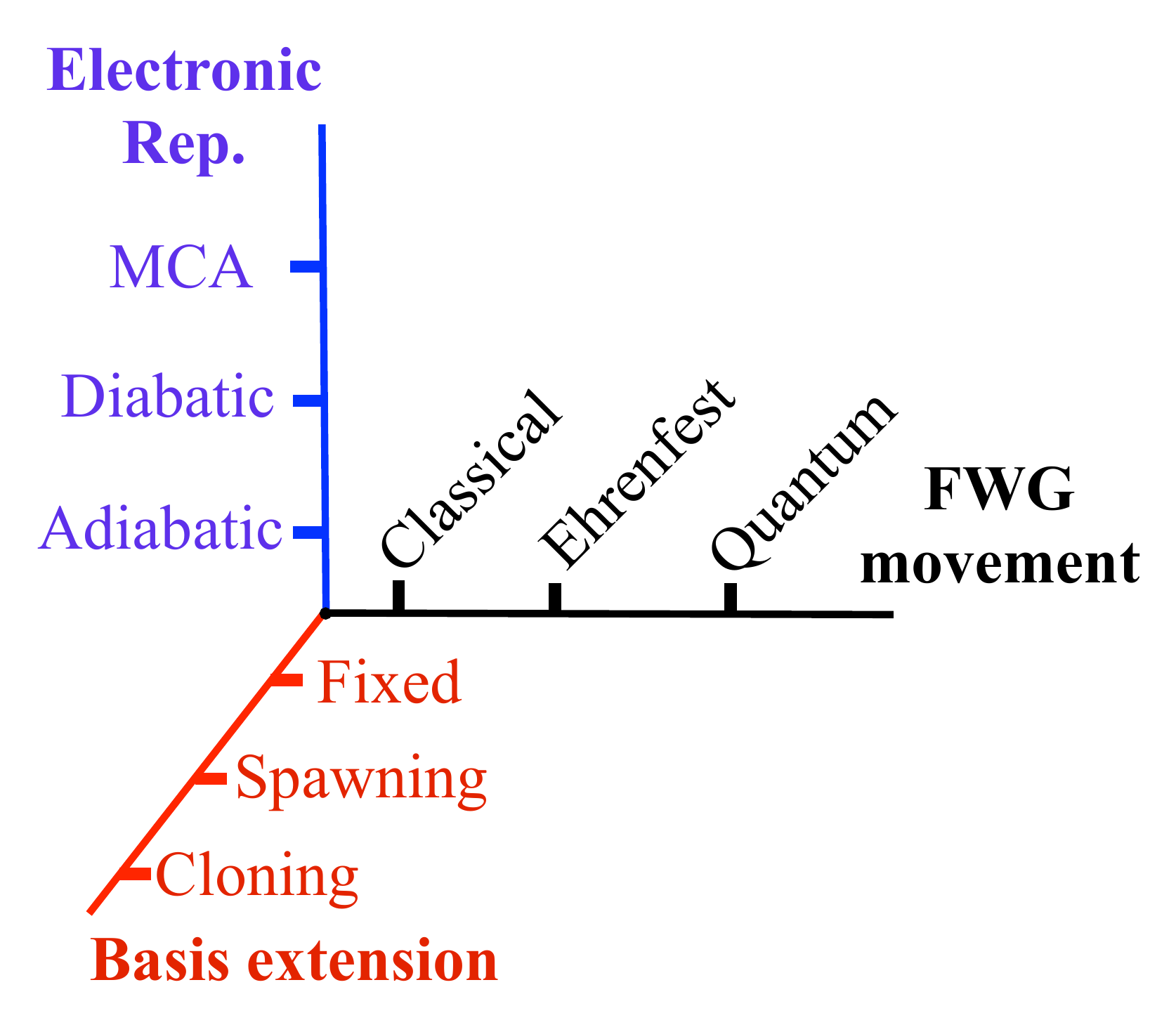}
  \caption{Independent components of FWG-based approaches.}
  \label{fig:sch}
\end{figure}

Recently, several comprehensive review articles on methods involving FWG propagation have been 
published.\cite{Makhov:2017/cc/200,Curchod:2018/cr,Richings:2015/irpc/269} 
What will be different in this work? (Even though it has not been intended as a review 
but rather a highlight of our efforts in developing FWG methods.) 
First, it will try to provide a unified perspective on several popular approaches. The key to this 
will be identifying few crucial issues that any method using \eq{eq:gwp} is facing and illustrating
how different schemes address these issues. Even from looking at \eq{eq:gwp} one can conclude that 
the accuracy of any approach will depend on the following choices: 1) nature of electronic states $\phi_s(r)$,
2) dynamics of individual FWGs, and 3) possibility for the basis set extension (see \fig{fig:sch}). 
The last aspect mainly concerns the nuclear basis (since the number of electronic states is usually predefined 
by the problem energy scale). Necessity to increase the number of FWGs stems from 
differences in complexity of $\Psi(r,R,0)$ and that at later times.  
Second, several fundamental issues related to the equivalence of different formulations, 
geometric phase treatment, and energy conservation will be highlighted. 
Third, we will discuss rigorous extensions of intuitive ideas of the basis extensions 
(e.g., spawning and cloning) and applying FWG methods 
for simulating open systems. 

Before starting our exposition, we would like to note that even though mixed quantum-classical (MQC) 
methods\cite{Tully:1998va,Tully:2012cz,Wang:2016bza} are very popular in the context of the nonadiabatic 
dynamics, and even though in many popular 
methods based on \eq{eq:gwp} FWGs move classically, we believe that comparison of FWG 
approaches to those of the MQC methodology can be misleading and we will try to avoid it. 
We would like to emphasize that FWG methods based on \eq{eq:gwp} are 
fully quantum, even though they have some restrictions on the form of the wavefunction, 
yet, both electronic $r$ and nuclear $R$ variables are present and completely independent 
from the time variable. On the other hand, MQC approaches turn the nuclear variables 
to classical trajectories, which are essentially functions of time. 
The main advantage of FWG methods with respect to the MQC ones 
is conceptually straightforward inclusion of all nuclear quantum effects (zero point energy, decoherence, etc.) 
because it is only a matter of a basis extension to build a nuclear wavefunction capable 
of describing the necessary effects.

Another popular method to model fully quantum nonadiabatic dynamics is \gls{MCTDH}.
\cite{Burghardt:2008iz,Burghardt:1999/jcp/2927,mey90:73,Wang:2003/jcp/1289} 
This method merely reduces the prefactor of the exponential scaling and allows one 
to consider larger systems than standard grid approaches\cite{Kosloff:1983va}. 
However, \gls{MCTDH} requires a global parametrization of PESs. 
One of the main reasons to prefer FWG formulation 
is avoidance of the PES parametrization and evaluating it on-the-fly.  
 
 The rest of this paper is organized as follows. 
First, we discuss the representations for electron and nuclear basis functions.
Second, we give the theoretical background on direct quantum dynamics and discuss the \gls{TDVP} formulation.
Then, we expose our extensions of popular methods for increase of the number of nuclear basis functions for variational dynamics.
Finally, we present a problem-free approach to open system dynamics using the \gls{NOSSE}.
 
\section{Representations}

\subsection{Frozen-width Gaussians}

One of the first introductions of FWGs to modeling dynamics of nuclear \glspl{DOF} was 
done by Heller.\cite{Heller:1975/jcp/1544,Heller:1976/jcp/63,Heller:1981/jcp/2923}
Among a few equivalent forms, FWGs in \eq{eq:gwp} are usually taken in the \gls{CS} form 
\bea
g_{k}(\mat R| \mat q_{k}(t),\mat p_{k}(t)) &=& \prod_{a=1}^{N_\mathrm{n}} 
\left(\frac{\omega_a}{\pi}\right)^{1/4} \nonumber\\
&\times& \prod_{\alpha=1}^\Dim\mathrm{e}^{-\frac{\omega_a}{2}(R_{a\alpha}-q_{ka\alpha})^2
+ip_{ka\alpha}(R_{a\alpha}-q_{ka\alpha})+ip_{ka\alpha}q_{ka\alpha}/2},\label{eq:cs}
\eea  
where $\omega_a$ controls the \gls{CS} width, $N_\mathrm{n}$ is the number of nuclear \glspl{DOF}, 
and $\Dim$ is the space dimensionality. This allows one to use the \gls{CS} algebra, 
positions of Gaussians centers and their momenta can be replaced by the complex variables $\{\mat z_{k}\}$ according to
\bea
z_{ka\alpha}(t) & = & \sqrt{\frac{\omega_a}{2}}q_{ka\alpha}(t) + \frac{i}{\sqrt{2\omega_a}}p_{ka\alpha}(t).
\eea
Using the complex variables $\{\mat z_{k}\}$, \glspl{CS} read
\bea
g_{k}(\mat R| \mat q_{k}(t),\mat p_{k}(t)) & = & g_{k}(\mat R|\mat z_{k}(t),\mat z_{k}^*(t)) = \prod_{a=1}^{N_\mathrm{n}} \left(\frac{\omega_a}{\pi}\right)^{1/4}G_{ka}(\vect R_a) \nonumber\\
G_{ka}(\vect R_a) & = & \prod_{\alpha=1}^\Dim\mathrm{e}^{-\frac{\omega_a}{2}(R_{a\alpha}-\sqrt{\frac{2}{\omega_a}}z_{ka\alpha})^2+iz_{ka\alpha}\Im(z_{ka\alpha})}.\label{eq:cs2}
\eea
With these definitions, we have the usual eigenvalue relation for \glspl{CS}
\bea
\left[ \sqrt{\frac{\omega_a}{2}}R_{a\alpha} + \frac{1}{\sqrt{2\omega_a}}\frac{\partial}{\partial R_{a\alpha}} \right] g_{k}(\mat R|\mat z_{k}(t),\mat z_{k}^*(t)) & = &
z_{ka\alpha}(t) g_{k}(\mat R|\mat z_{k}(t),\mat z_{k}^*(t)).
\eea
The \gls{CS} choice for the form of FWGs does not only simplify derivations but also removes some of the 
numerical issues that plagued other choices.\cite{Saita:2012/jcp/22A506} One of the main contributions 
to the success of the \gls{CS} form is the classical action represented in the phase part
$i\sum_{a,\alpha} p_{ka\alpha}q_{ka\alpha}/2$ in \eq{eq:cs}.
Another FWG choice used in 
\gls{vMCG}\cite{Worth:2008/MP/2077} had an extra time-dependent variable that 
represented this phase and required an additional time-integration to obtain its 
time-dependence. This extra time-integration was extremely unstable because of highly oscillatory 
nature of the phase. The classical analogue represented in the CS form captures the majority 
of the true quantum phase and does not require an additional time-integration.

The nuclear variables $R$ can be either internal \glspl{DOF} (e.g. normal modes) or Cartesian coordinates.
In the latter case, optimal $\omega_a$'s obtained for individual atoms can be used.\cite{Thompson:2010/cp/70}

\subsection{Electronic wavefunctions}
\label{sec:rep}

There are two main choices for the electronic states in \eq{eq:gwp}, adiabatic and diabatic wavefunctions. 
In this subsection we summarize pros and cons for both representations 
and highlight a newly developed \gls{MCA} representation 
that combines the best properties of the conventional representations and has been introduced 
recently in the context of the FWG dynamics.\cite{Joubert:2017/jpcl/452} 

\paragraph{Adiabatic representation:} The electron-nuclear wavefunction without 
using a specific FWG nuclear representation can be written as the \gls{BH} expansion
\bea\label{eq:adBH}
\Psi(r,R,t) = \sum_s \phi_s(r|R)\chi_s(R,t),
\eea 
where $\phi_s(r|R)$ electronic eigenstates of the electronic Hamiltonian 
$H_e(r|R)\phi_s(r|R) = E_s(R)\phi_s(r|R)$, and $\chi_s(R,t)$ are the nuclear time-dependent wavefunctions. 
To obtain equations for $\chi_s(R,t)$, the total TDSE is projected onto the space of the electronic 
wavefunctions and the integration over the electronic \gls{DOF} is
done
\bea\label{eq:eomad}
\sum_{s'}[(T_n + E_s(R))\delta_{ss'} + \Lambda_{ss'} ]\chi_{s'}(R,t) = i\partial_t \chi_s(R,t)
\eea
Due to the parametric dependence of the adiabatic states on the nuclear \gls{DOF}, this representation generates \glspl{NAC}
\bea
\Lambda_{ss'} & = & -\sum_{a\alpha}\frac{1}{2M_a} \braket{\phi_s(R)}{\frac{\partial^2 \phi_{s'}(R)}{\partial R_{a\alpha}^2}}_r -\sum_{a\alpha}\frac{1}{M_a} \braket{\phi_s(R)}{\frac{\partial \phi_{s'}(R)}{\partial R_{a\alpha}}}_r\frac{\partial}{\partial R_{a\alpha}} 
\eea
Among these terms, the \gls{DBOC}, 
\bea
\braket{\phi_s(R)}{\frac{\partial^2 \phi_s(R)}{\partial R_{a\alpha}^2}}_r \sim \frac{1}{(E_s(R)-E_{s\pm 1}(R))^{2}}
\eea
diverges at commonly encountered \glspl{CI}~\cite{Yarkony:1996/rmp/985,Migani:2004/271} 
and makes it impossible to integrate \eq{eq:eomad}.~\cite{meek:2016c}

An additional problem arising due to \glspl{CI} is appearance of nontrivial Berry 
or \gls{GP} in the electronic wavefunctions of states involved in the 
\glspl{CI}.~\cite{LonguetHigg:1958/rspa/1,Berry:1984/rspa/45,Mead:1992/rmp/51} 
This phase makes the electronic wavefunction change their signs if one evolves them continuously 
around the CI seam. Thus, the electronic wavefunctions of states experiencing 
the CI are double-valued with respect to the nuclear variables $R$.
This extra feature profoundly affects nuclear dynamics, both for cases when the system evolves 
on a single PES involved in the \gls{CI} and when the system participates in the interstate 
transitions through the \gls{CI}.\cite{Ryabinkin:2017/acr/1785} Ignoring the GP can lead 
to completely inadequate nuclear dynamics.\cite{Ryabinkin:2013/prl/220406,Loic:2013/jcp/234103,Guo:2016/jacs,Joubert:2017/cc/7365,Xie:2018/jacs/1986,Li:2017/jcp/064106,Henshaw:2018ck}
Accounting for the \gls{GP} can be done by introducing a resolution of the identity 
$1= \exp(i\theta(R))\exp(-i\theta(R))$ to the \gls{BH} expansion (\eq{eq:adBH}), 
where $\exp(i\theta(R))$ is a double-valued 
function that compensates double-valuedness of both electronic and nuclear 
wavefunctions.\cite{Mead:1979/jcp/2284}
However, constructing $\theta(R)$ for the on-the-fly approaches using FWGs can be challenging 
because $\theta(R)$ should encode global topological properties of the CI seam positions. 

One can see that the origin of all the problems stems from the $R$-dependence of electronic 
wavefunctions $\phi_s(r|R)$ and the nuclear kinetic energy operator terms $\Lambda_{ss'}$ 
originating from this dependence. On the other hand, the $R$-dependence in $\phi_s(r|R)$ 
allows one to have a compact electronic representation because 
calculating the electronic states in a single point (crude adiabatic representation)
would require a long expansion for the total wavefunction. A natural solution is to minimize the negative
impact of the nuclear dependence in the electronic wavefunctions, 
and it leads to a so-called diabatic representation. 

\paragraph{Diabatic representation:}
Changing from adiabatic to diabatic representation removes the \glspl{NAC} and the 
\gls{GP} and thus resolves both problems~\cite{Richings:2015/irpc/269,Richings:2017/cpl/606,Makhov:2017/cc/200,Meek:2016/jcp/184103}
\bea\label{eq:diBH}
\Psi(r,R,t) = \sum_s \phi_s(r) \tilde{\chi}_s(R,t),
\eea
where
\bea\label{eq:a2d}
\phi_s(r) = \sum_{s'}U_{ss'}(R)\phi_{s'} (r|R).
\eea
This leads to the analogous equations for nuclear wavefunctions 
\bea
\sum_{s'}[T_N\delta_{ss'} + V_{ss'} ]\tilde{\chi}_{s'}(R,t) = i\partial_t \tilde{\chi}_s(R,t),
\eea
where $V_{ss'} = \braket{\phi_s}{H_e\vert\phi_{s'}}_r$. 

Unfortunately, the diabatization transformation (\eq{eq:a2d}) is generally 
exact only in a complete set of electronic states, 
and is approximate otherwise. In the latter case, the transformation in \eq{eq:a2d} is known as quasi-diabatization.
Quasi-diabatizations are employed to remove the largest, singular at \glspl{CI}, part of the \glspl{NAC}.
Quasi-diabatization technics for on-the-fly (direct) dynamics include regularized diabatization,~\cite{Koppel:2006/mp/1069,Worth:2008/MP/2077,Richings:2015/irpc/269} local integration of the \glspl{NAC},~\cite{Richings:2017/cpl/606} or time-dependent quasi-diabatization.~\cite{Meek:2016/jcp/184103}
While the remaining part of the \glspl{NAC} is considered to be negligible, the error introduced is difficult to control.
Furthermore, for direct dynamics, such a quasi-diabatization can only be done in some regions of the nuclear space (e.g. at a known \gls{CI} position or at FWG centers), which can leave singularities in 
 other regions of the nuclear space. The recently introduced \gls{DGAS} basis solves this problem by applying a local 
diabatization at each FWG center.~\cite{Meek:2016/jcp/184103}

\paragraph{MCA representation:}
All problems of the global diabatic and adiabatic representations are avoided 
by simply replacing the dependence of the electronic states on the nuclear \gls{DOF} by 
their dependence on the FWGs' centers, $q_k(t)$,
\bea\label{eq:MCA_Psi}
\Psi(r,R,t) = \sum_{k,s} C_{ks}(t) g_k(R|q_k(t),p_k(t)) \phi_s(r|q_k(t)).
\eea
The resulting electronic states $\phi_s(r|q_k(t))$ 
are adiabatic only at $\mat q_k$ but used for any other nuclear geometries 
and are commonly referred to as crude adiabatic states.~\cite{Longuet:1961/as/429,Ballhausen:1972/arpc/15}
\glslocalreset{MCA}
Since these crude adiabatic states are attached to the moving Gaussians, we refer to the expansion in \eq{eq:MCA_Psi} as the \gls{MCA} representation,~\cite{Joubert:2017/jpcl/452} 
also known as time-dependent diabatic basis.~\cite{Makhov:2017/cc/200}
In fact, the MCA representation is a rigorous diabatic representation because the electronic wavefunctions 
do not have any $R$-dependence. It becomes very important to distinguish the nuclear variables $R$ 
and the FWGs' centers $q_k(t)$. As $\phi_s(r|q_k(t))$ do not depend on nuclear \gls{DOF}, 
the nuclear kinetic energy operator has no effect on the \gls{MCA} states, so that \glspl{NAC} are exactly zero.

It is also possible to show a precise mechanism of the \gls{MCA} solution of the GP problem. 
The \gls{MCA} representation can be obtained starting from an exact complete diabatic representation $\{\phi_s\}$ by applying rotations $\mat Q(q_k)$ to a set of adiabatic states at the Gaussians' centers 
\bea
\op H_\mathrm{e} [\mat q_k] \sum_{u} \ket{\phi_{u}} Q_{us}(q_k) & = & \op H_\mathrm{e} [\mat q_k] \ket{\phi_s(q_k)} = E_s (\mat q_k) \ket{\phi_s(q_k)}.
\eea
Applying this transformation to the total wavefunction in the diabatic representation 
gives the \gls{MCA} representation
\bea\label{eq:transf_diab_MCA}
\ket{\Psi} & = & \sum_{ku} \tilde{C}_{ku} \ket{g_k} \ket{\phi_u} \nonumber\\
& = & \sum_{ks} \left(\sum_{u} \tilde{C}_{ku} Q_{us}(q_k)\right) \left(\sum_{v} \ket{\phi_v} Q_{vs}(q_k)\right) \ket{g_k} \nonumber\\
& = & \sum_{ks} {C}_{ks}(q_k) \ket{g_k} \ket{\phi_s(q_k)},
\eea
where we used orthogonality of the rotation matrix $\mat Q(q_k) {\mat Q(q_k)}^T=\mat 1$, while 
the last equality is \eq{eq:MCA_Psi}.
For \glspl{CI}, $\mat Q(q_k)$ is double-valued due to appearance of \gls{GP} 
along any path followed by $\mat q_k$ encircling the \gls{CI}.
Thus, the \gls{MCA} states are double-valued as functions of $\mat q_k$.
\Eq{eq:transf_diab_MCA} shows that the coefficients ${C}_{ks}(q_k)$ are also double-valued in the parameter space and compensate for the double-valuedness of the \gls{MCA} states.
By construction, the \gls{MCA} representation provides double-valuedness to nuclear wavepackets 
$\sum_{k}{C}_{ks}(q_k)\ket{g_k}$ and the \gls{GP} is naturally included in the approach.

\subsection{Multi-set and single-set}

Upon deciding on the form of FWGs and the electronic state representation, there are still two choices left in 
combining the electronic and nuclear functions in linear combination of \eq{eq:gwp}. The key difference 
between these choices is whether FWG positions and momenta depend on the electronic state 
multiplying FWGs, {\it multi-set} 
\bea\label{eq:ms}
\Psi(r,R,t) = \sum_{k,s} C_{ks}(t) g_k(R|q_k^{(s)}(t),p_k^{(s)}(t)) \tilde{\phi}_s(r)
\eea
or not, {\it single-set} 
\bea\label{eq:ss}
\Psi(r,R,t) = \sum_{k,s} C_{ks}(t) g_k(R|q_k(t),p_k(t)) \tilde{\phi}_s(r),
\eea
where $\tilde{\phi}_s(r)$ can be electronic states of any representation discussed above. Both sets converge to 
the same limit, but they start from different points. 
\begin{figure}
  \centering
  \includegraphics[width=0.5\textwidth]{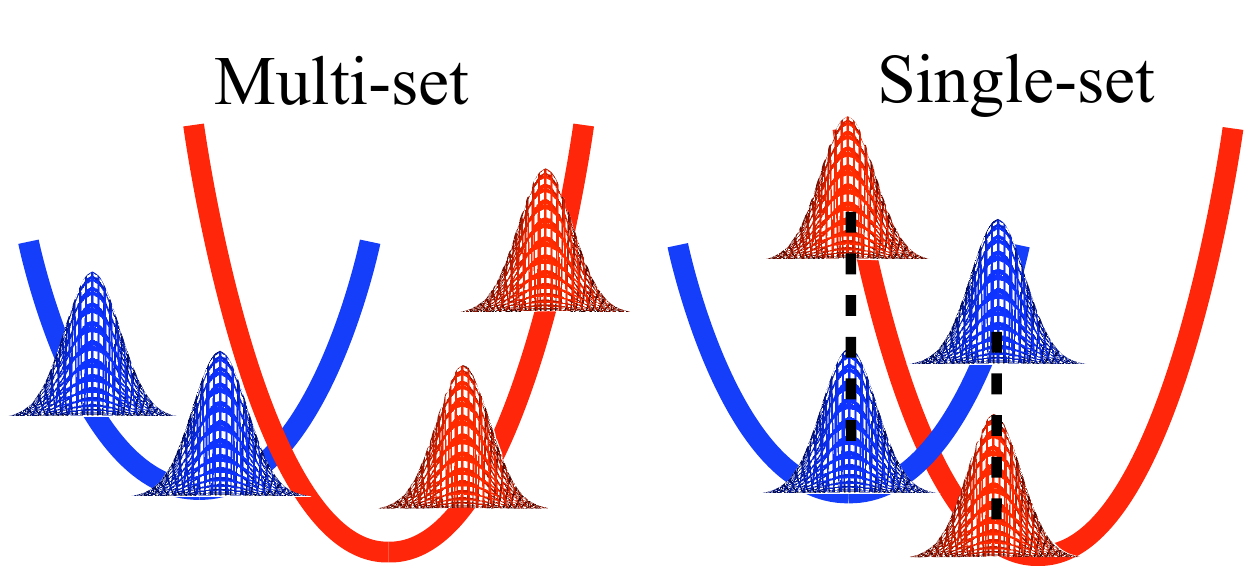}
  \caption{Illustration of multi-set and single-set FWGs.}
  \label{fig:ssms}
\end{figure}
Multi-set is more adapted toward nuclear dynamics that 
is very different on different electronic states (\fig{fig:ssms} left), an alternative form of \eq{eq:ms} illustrating this view is 
\bea\label{eq:ms2}
\Psi(r,R,t) = \sum_{s} \tilde{\phi}_s(r) \left[\sum_k C_{ks}(t) g_k(R|q_k^{(s)}(t),p_k^{(s)}(t))\right], 
\eea
here, every electronic wavefunction has its own linear combination of FWGs.  
In contrast, the starting point of the single-set is similar dynamics of each FWG on multiple electronic 
states (\fig{fig:ssms} right), or alternatively, it is a motion of a stack of FWGs with the 
same positions and momenta but 
different amplitudes. Therefore, each FWG in single-set can be thought as multiplied by its own time-dependent 
superposition of electronic states 
\bea\label{eq:ss2}
\Psi(r,R,t) = \sum_{k} g_k(R|q_k(t),p_k(t))\left[\sum_s C_{ks}(t) \tilde{\phi}_s(r)\right].
\eea
Both sets have been used before, the main advantage of multi-set is perfect description of nuclear decoherence
effects while making it difficult for FWGs on different electronic states to interact because of vanishing nuclear 
space overlap. The opposite is true for the single-set where the main advantage is for replicas of a FWG in 
the same stack to exchange the population readily because of the perfect overlap. 
On the other hand, description of nuclear decoherence in the single-set requires increasing numbers of 
FWGs.   
  
\section{Equations of motion}
\label{sec:EOM}

A general approach to obtain differential equations for time-dependence of FWG parameters  
involves application of the \gls{TDVP}. In this section we will illustrate caveats related to 
various forms of the \gls{TDVP}, its partial application, and energy and norm conservation
problems.

\subsection{Various forms of the time-dependent variational principle}

Three different forms of the TDVP can be found, first, the Dirac-Frenkel (DF) 
TDVP\cite{Dirac:1930/PCPS/376,Book/Frenkel:1934}
\bea\label{eq:DF}
\braket{\delta\Psi}{H-i\partial_t|\Psi} & = & 0,
\eea 
which is based on imposing orthogonality of all possible variations allowed by the parametrization, $\delta\Psi$, 
to the error vector in the TDSE due to the wavefunction parametrization, $\ket{(H-i\partial_t)\Psi}$. 
This orthogonality ensures the optimal choice of parameters' time-dependence, since the 
residual error cannot be reduced further by any variation of the parameters. 
Second, the McLachlan TDVP,\cite{McLachlan:1964/mp/39} 
which is rooted in minimization of the error vector norm 
$||(H-i\partial_t)\ket{\Psi}||$, results in\cite{Broeckhove:1988wo} 
\bea\label{eq:ML}
\Im\braket{\delta\Psi}{H-i\partial_t|\Psi} & = & 0.
\eea 
Third, the stationary action principle, here the quantum action defined as\cite{Broeckhove:1988wo}
\bea
\mathcal{S}[\Psi] = \int_{t_0}^{t_1} \braket{\Psi}{H-i\partial_t|\Psi} dt
\eea
is varied with respect to wavefunction parameters' as functions of time. 
The condition $\delta \mathcal{S}[\Psi] =0$ is equivalent to the equation\cite{Broeckhove:1988wo} 
\bea\label{eq:AVP}
\Re\braket{\delta\Psi}{H-i\partial_t|\Psi} & = & 0.
\eea 
Yet, all these versions of the TDVP are found to give the same \glspl{EOM} if the wavefunction parametrization 
is analytic in the complex variable sense (function $f$ of complex variable $z$ is 
analytic if $\partial f(z)/\partial z^* = 0$).\cite{Broeckhove:1988wo} 
Functions based on CSs are generally non-analytic because of $z_{ka\alpha}^*$ dependencies in CSs (\eq{eq:cs2}). 
Non-analyticity of electron-nuclear basis products
can be decomposed in two components originating from nuclear and electronic parts
\bea\label{eq:nan}
\ket{\frac{\partial  (g_{k} \phi_s)}{\partial z_{ka\alpha}^*}} & = & \ket{\frac{\partial g_{k}}{\partial z_{ka\alpha}^*}\phi_s} + \ket{g_{k}\frac{\partial \phi_s}{\partial z_{ka\alpha}^*}}.
\eea
The global adiabatic and diabatic representations do not have the second electronic term in \eq{eq:nan}. 
For these representations even the non-analyticity stemming from individual CSs do not affect the \glspl{EOM}
because terms corresponding to $z_{ka\alpha}^*$ derivatives vanish. Thus, all three forms of the TDVP 
give the same \glspl{EOM} for the global adiabatic and diabatic representations. 
One of the most well-known approaches derived 
in the diabatic representation using the DF TDVP is the \gls{vMCG} method.~\cite{Richings:2015/irpc/269} 
 
 
Similar to the other representations, for MCA, non-analyticity of individual CSs (the first term in \eq{eq:nan}) 
does not contribute to \glspl{EOM}, however, the electronic wavefunctions depend on 
$q_{ka}\sim z_{ka\alpha}+ z_{ka\alpha}^*$, and therefore, due to this non-analyticity all TDVP
forms give different \glspl{EOM} for the MCA representation. 
Among the three TDVP versions, only the stationary action principle is 
proven to lead to \glspl{EOM} necessarily conserving the system energy,\cite{Kan:1981/pra/2831} 
and thus, we illustrate the MCA \glspl{EOM} based on this principle.

We will use the MCA representation with the single-set expansion, 
$\ket{\Psi} = \sum_{ks} C_{ks}\ket{\varphi_{ks}}$ where the electron-nuclear basis functions $\{\varphi_{ks}\}$ are 
\bea\label{eq:Psi_adiab}
\varphi_{ks}(\mat R,\mat r|\mat z_{k}(t),\mat z_{k}^*(t)) & = & g_{k}(\mat R|\mat z_{k}(t),\mat z_{k}^*(t)) 
\phi_s(\mat r|\mat z_{k}(t),\mat z_{k}^*(t)).
\eea
In order to simplify equations we will use the short-hand notation, $\phi_s(\mat r|\mat z_{k}(t),\mat z_{k}^*(t))\equiv\ket{\phi_s^k}$, and will omit the time and nuclear coordinates.
Using \eq{eq:AVP} we obtain the system of coupled equations
\bea
\partial_t{\mat C} & = & \mat S^{-1}(-i\mat H-\mat\gamma)\mat C, \label{eq:Cdot} \\
\mat B \partial_t{\mat z} + \mat A \partial_t{\mat z}^* & = & \mat Y + \overline{\mat Y}, \label{eq:zdot}
\eea
where matrices and vectors are defined as
\bea
S_{kl,ss'} & = & \braket{\varphi_{ks}}{\varphi_{ls'}}, \label{eq:S}\\
H_{kl,ss'} & = & \braOket{\varphi_{ks}}{\op H}{\varphi_{ls'}}, \label{eq:H}\\
\gamma_{kl,ss'} & = & \braket{\varphi_{ks}}{\partial_t\varphi_{ls'}}, \label{eq:gamma}
\eea
\bea
Y_{ka\alpha} & = & i\sum_{lss'} C_{ks}^*\braOket{\frac{\partial\varphi_{ks}}{\partial z_{ka\alpha}}}{\op 1-\op{\mathcal{P}}}{\op H\varphi_{ls'}}C_{ls'}, \label{eq:Y}\\
\overline Y_{ka\alpha} & = & i\sum_{lss'} C_{ls'}^*\braOket{\op H\varphi_{ls'}}{\op 1-\op{\mathcal{P}}}{\frac{\partial\varphi_{ks}}{\partial z_{ka\alpha}^*}}C_{ks}, \label{eq:Ybar}\\
A_{kl,ab,\alpha\beta} & = & \sum_{ss'} \Bigg[ C_{ls'}^*\braOket{\frac{\partial\varphi_{ls'}}{\partial z_{lb\beta}}}{\op 1-\op{\mathcal{P}}}{\frac{\partial\varphi_{ks}}{\partial z_{ka\alpha}^*}}C_{ks} \nonumber\\
&&\hspace{0.cm}- C_{ks}^*\braOket{\frac{\partial\varphi_{ks}}{\partial z_{ka\alpha}}}{\op 1-\op{\mathcal{P}}}{\frac{\partial\varphi_{ls'}}{\partial z_{lb\beta}^*}}C_{ls'} \Bigg], \label{eq:A}\\   
B_{kl,ab,\alpha\beta} & = & \sum_{ss'} \Bigg[ C_{ls'}^*\braOket{\frac{\partial\varphi_{ls'}}{\partial z_{lb\beta}^*}}{\op 1-\op{\mathcal{P}}}{\frac{\partial\varphi_{ks}}{\partial z_{ka\alpha}^*}}C_{ks} \nonumber\\
&&\hspace{0.cm}- C_{ks}^*\braOket{\frac{\partial\varphi_{ks}}{\partial z_{ka\alpha}}}{\op 1-\op{\mathcal{P}}}{\frac{\partial\varphi_{ls'}}{\partial z_{lb\beta}}}C_{ls'} \Bigg]. \label{eq:B}   
\eea
Equations~(\ref{eq:Y}-\ref{eq:B}) involve the projector on the non-orthogonal basis 
\bea\label{eq:Proj}
\op{\mathcal{P}} & = & \sum_{kl,ss'} \ket{\varphi_{ks}}[\mat S^{-1}]_{kl,ss'}\bra{\varphi_{ls'}}.
\eea
Here, it is easy to see that the non-analyticity from the CSs derivatives vanish, the 
corresponding terms appearing in Equations~(\ref{eq:Y}-\ref{eq:B}) will contain projections 
of the derivatives
\bea
(\op 1-\op{\mathcal{P}})\ket{\frac{\partial g_{k}}{\partial z_{ka\alpha}^*}\phi_s^k} & = & -(\op 1-\op{\mathcal{P}})\frac{z_{ka\alpha}}{2}\ket{\varphi_{ks}}\equiv 0,
\eea
where we used the CS definition to obtain the explicit form of the derivatives. These derivative terms vanish 
because projector $\op{\mathcal{P}}$ is equivalent to the identity operator for the basis $\{\varphi_{ks}\}$. 

Equation~(\ref{eq:zdot}) can be solved by combining it with its complex conjugate to the system of equations
\bea\label{eq:zdot2}
\begin{pmatrix}
\mat B         &  \mat A   \\
\mat A^\dagger & -\mat B^T
\end{pmatrix}
\begin{pmatrix}
\partial_t{\mat z} \\
\partial_t{\mat z}^*
\end{pmatrix}
& = &
\begin{pmatrix}
{\mat Y} + \overline{\mat Y} \\
-{\mat Y}^* - \overline{\mat Y}^*
\end{pmatrix},
\eea
and then solving the system by block matrix inversion
\bea\label{eq:zdotsolve}
\begin{pmatrix}
\partial_t{\mat z} \\
\partial_t{\mat z}^*
\end{pmatrix}
& = &
\begin{pmatrix}
\mat \beta          &  \mat \alpha  \\
\mat \alpha^\dagger & -\mat \beta^T
\end{pmatrix}
\begin{pmatrix}
{\mat Y} + \overline{\mat Y} \\
-{\mat Y}^* - \overline{\mat Y}^*
\end{pmatrix},
\eea
where 
\bea
\mat \beta & = & [ \mat B + \mat A (\mat B^T)^{-1}\mat A^\dagger]^{-1}, \\
\mat \alpha & = & \mat B^{-1}\mat A \mat \beta^T = \mat \beta \mat A (\mat B^T)^{-1}.
\eea
The obtained equations is somewhat similar to those in other methods using FWGs and 
the TDVP such as \gls{vMCG} and G-MCTDH, but there are clear differences stemming 
from the MCA representation. 

\subsection{Conservation of norm and energy}

Equation~(\ref{eq:Cdot}) have the property that the norm of the wavefunction is conserved by construction
\bea\label{eq:norm_conservation}
\partial_t\braket{\Psi}{\Psi} & = & 2\Re\braket{\Psi}{\partial_t\Psi} \nonumber\\
& = & 2\Re\left( {\mat C}^\dagger\mat S\partial_t{\mat C} + \mat C^\dagger{\mat \gamma}\mat C \right) \nonumber\\
& = & 2\Im\left( {\mat C}^\dagger\mat H\mat C \right) = 0,
\eea
where we used hermiticity of $\mat H$. 
The system energy is also conserved by construction for a time-independent Hamiltonian, as it can be verified by using Eqs.~(\ref{eq:Cdot},\ref{eq:Y},\ref{eq:Ybar},\ref{eq:zdotsolve}):
\bea\label{eq:energy_conservation}
\partial_t E & = & 2\Re\braOket{\Psi}{\op H}{\partial_t\Psi} \nonumber\\
& = & 2\Re\left[ i\overline{\mat Y}^\dagger\partial_t{\mat z} - i\mat Y^T\partial_t{\mat z}^* \right] \nonumber\\
& = & 2\Im\left[ \begin{pmatrix} \overline{\mat Y} \\ -\mat Y^* \end{pmatrix}^\dagger \begin{pmatrix} \mat \beta & \mat \alpha \\ \mat \alpha^\dagger & -\mat \beta^T  \end{pmatrix} \begin{pmatrix} \mat Y + \overline{\mat Y} \\ -\mat Y^* - \overline{\mat Y}^* \end{pmatrix} \right], \nonumber\\
& = & -2\Im\left[ \begin{pmatrix} \overline{\mat Y}^* \\ -\mat Y \end{pmatrix}^T \begin{pmatrix} \mat \alpha & \mat \beta \\ -\mat \beta^T & \mat \alpha^\dagger  \end{pmatrix} \begin{pmatrix} \overline{\mat Y}^* \\ -\mat Y \end{pmatrix} \right] = 0.
\eea
In the last two equalities we used that $\mat \beta$ and $\mat \alpha$ are hermitian and antisymmetric, respectively.

\subsection{Classically moving Gaussians}

The TDVP equations for $z$ and $z^*$ (or effectively $p$ and $q$) 
are the most time-consuming because of the dimensionality of the 
involved matrices. One can apply the TDVP to FWG expansions to obtain \glspl{EOM} only for 
the amplitudes $C_{ks}$, while \glspl{EOM} for the FWG evolution can be chosen to be classical 
(or Ehrenfest). This reduced approach still 
gives rise to quantum formalism because both electronic and nuclear wavefunctions are present.
It has been used in the \gls{MCE} method of Shalashilin~\cite{Saita:2012/jcp/22A506,Makhov:2017/cc/200} 
(single-set basis) and the \gls{AIMS} method of Martinez~\cite{BenNun:2002tx,Yang:2009ja,Curchod:2018/cr}
(multi-set basis). These approaches have the advantage that each FWG propagation can be done 
independently (in parallel) from others, so that, later, the information on individual trajectories can be used 
for solving \eq{eq:Cdot} and obtaining the amplitudes $C_{ks}$.

However, using classical \glspl{EOM} for FWGs affects two properties, which are intertwined in this case: 
the convergence with respect to the number of nuclear basis functions and the energy conservation. 
Classical motion of FWGs increases requirements for the number of basis functions in general. 
It is related to quantum forces appearing in the full TDVP formalism that make FWG movement
more optimal.\cite{Worth:2008/MP/2077} Also, formalisms using classical FWG motion do not
conserve the total energy.\cite{Habershon:2012/jcp/014109} The energy conservation problem 
can be reduced by increasing the basis set because at the complete basis set limit the motion of FWGs
does not affect the formalism. On the other hand, the energy conservation becomes especially problematic 
when two or more FWGs overlap significantly in the course of their dynamics. The energy conserving 
FWG motion is predicted by the full TDVP, whereas deviations from it based on classical dynamics 
can be considered as interference of an external force. Clearly, the total energy is not generally 
conserved in this case. In contrast, if FWGs do not overlap significantly, TDVP FWG motion reduces 
to their classical motion in accordance with the Ehrenfest theorem, and the energy is conserved. 
The latter scenario is more probable for general, large dimensional systems, where nuclear decoherence
is usually fast. Nevertheless, due to advantages provided by independent FWG evolution schemes, 
the work on developing a FWG-based method with uncoupled trajectories that would conserve the 
energy is highly desirable. 

\subsection{Integral evaluation} 

Solving Eqs.~(\ref{eq:Cdot}) and (\ref{eq:zdot2}) require matrix elements involving 
integration with respect to both electronic and nuclear coordinates. 
For the global adiabatic and diabatic representations, the 
TDVP is effectively applied to Eqs.~(\ref{eq:adBH}) and (\ref{eq:diBH}) 
because the electronic states in these representations 
do not contain parameters subject to the optimization. 
In solving these equations, the electron-nuclear integration is done in two steps, first,
the electronic variables are integrated in electronic structure programs, 
and then nuclear dependent quantities are integrated with FWGs. 
In relation to the nuclear integration, 
the diabatic models are convenient because all integrals with nuclear FWGs 
are analytic due to polynomic form of the $V_{ss'}(R)$ functions. 
The adiabatic representation is more challenging due to less smooth behavior of PESs 
and divergencies of NACs near CI seams. 
Furthermore, adiabatic and diabatic states dependence on nuclear coordinates
is generally not known so that one must resort to local approximate models for 
the nuclear integration. 
Locality of FWGs helps for introducing approximations 
into matrix elements arising in solving \eq{eq:eomad}. Let us consider matrix element 
$\braket{g_k(q_k)\vert A(R)}{g_l(q_l)}_R$ appearing in computational schemes based on \eq{eq:eomad} 
and expanding $\chi_s(R,t)$ as a linear combination of FWGs, here, $A(R)$ is either PES or NAC. 
Note, that a product of two FWGs, $g_k(R|q_k)$ and $g_l(R|q_l)$, is again a FWG centered in 
between the centers of the two FWGs, $q_c = (q_k+q_l)/2$, 
this relation is also known as the Gaussian Product Rule.  
A typical estimation of the integral of interest involves a Taylor series expansion of $A(R)$ at 
$q_c$\cite{Curchod:2018/cr} 
\bea\notag
\braket{g_k(q_k)\vert A(R)}{g_l(q_l)}_R &\approx& A(q_c)\braket{g_k(q_k)}{g_l(q_l)}_R+
\sum_{a\alpha}\frac{\partial A(R)}{\partial R_{a\alpha}}\Bigg\vert_{R=q_c} \\ \notag
&&\times\braket{g_k(q_k) \vert (R_{a\alpha}-q_{c,a\alpha})}{g_l(q_l)}_R+
\frac{1}{2}\sum_{a\alpha,b\beta}\frac{\partial^2 A(R)}{\partial R_{a\alpha}\partial R_{b\beta}}\Bigg\vert_{R=q_c} \\
&&\times\braket{g_k(q_k)\vert (R_{a\alpha}-q_{c,a\alpha})(R_{b\beta}-q_{c,b\beta})}{g_l(q_l)}_R,
\eea 
where the integrals of FWGs with polynoms of $R$ (Gaussian moments) are analytical, but calculating
the $A(R)$ derivatives increases the computational cost. 

One of the greatest advantages of the MCA representation is that typical approximations for 
PESs and NACs can be avoided completely, even though the MCA representation 
comes from solving the electronic structure problem like the adiabatic representation.
The key element for avoiding approximations is a different partitioning of the total molecular Hamiltonian,
$H = h_e(\mat r)+h_n(\mat R)+V_{en}(|\mat r-\mat R|)$, here by grouping all electron and nuclear 
variables in $h_e(\mat r)$ and $h_n(\mat R)$ we expose the only electron-nuclear Coulomb coupling term, $V_{en}$.  
This partitioning does not even introduce PESs as intermediate quantities. Instead, matrix elements
such as in \eq{eq:H} can be evaluated numerically exactly
\bea
\braOket{\varphi_{ks}}{\op H}{\varphi_{ls'}} &=& \braOket{\phi_{ks}}{\op h_e}{\phi_{ls'}}_r\braket{g_k}{g_l}_R+
\braOket{g_{k}}{\op h_n}{g_{l}}_R\braket{\phi_s}{\phi_{s'}}_r + \braOket{\varphi_{ks}}{\op V_{en}}{\varphi_{ls'}}. \nonumber\\
\eea  
Note that both electronic and nuclear basis functions are FWGs at a single particle level, therefore, 
effective Gaussian integration techniques\cite{Book/Helgaker:2000} can be used.
The other matrix elements (\eqs{eq:S}-(\ref{eq:B})) also benefit from the new Hamiltonian partitioning 
and the MCA factorization of electronic and nuclear variables.\cite{Joubert:2018/JCP}
Therefore, the \gls{MCA} representation provides a framework that is free of any approximation other than finite number of the basis
functions.


\section{Nuclear basis set extensions}

\begin{figure}
  \centering
  \includegraphics[width=0.5\textwidth]{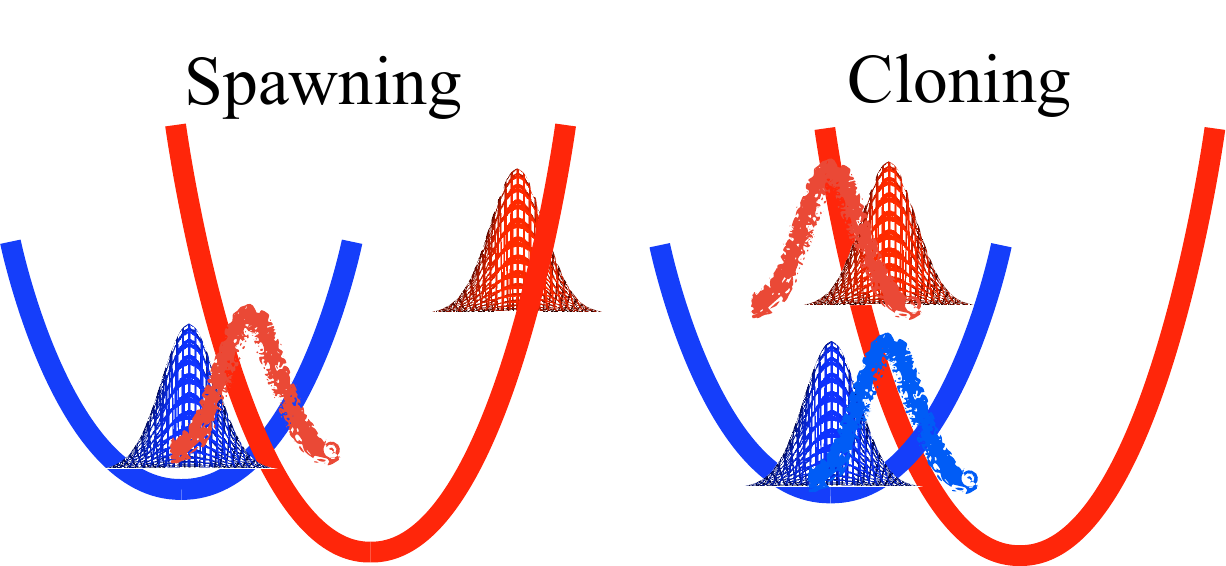}
  \caption{Illustration of spawning and cloning basis set extensions.}
  \label{fig:spcl}
\end{figure}

One of the main limitations of the FWG based schemes is a finite number of basis functions. 
Usually, the system wavefunction increases its complexity with time. 
Any finite number of basis functions will not be able to keep up with this increase. 
An intuitive solution is to devise an algorithm that cautiously increases the number of basis 
functions along the course of dynamics. Depending on a particular form of the used set, multi- or single-set,
there were two ideas put forward: spawning\cite{BenNun:2002tx} and cloning\cite{Makhov:2014/jcp/054110} (\fig{fig:spcl}).
In spawning for multi-set, the main problem of a finite basis is that dynamics 
of FWGs on different electronic states are governed by different forces. This leads to 
quick nuclear decoherence and thus decay of nuclear overlaps (in \eq{eq:S}) 
between FWGs on different surfaces.
Without the overlap the transition of the population between electronic states is impossible,
therefore, even if a FWG approaches a region of strong coupling (\fig{fig:ssms} left) it cannot 
pass any population unless there is an overlapping FWG on the other electronic state. 
To correct for this inadequacy of a finite basis set expansion, the spawning procedure would 
create a new FWG with an initial zero amplitude 
so that needed population transfer can occur (\fig{fig:spcl} left). 

In the single-set basis, the population transfer can always occur due to the perfect overlap between 
replicas of any FWGs in the same stack (\fig{fig:ssms} right). 
The main problem of the single-set finite basis representation is ``over-coherence''. 
In other words, a FWG stack always moves influenced by an averaged electronic state. 
To correct for ``over-coherence'', it is sensible to split stacks of FWGs when 
individual components experience forces which are very different. To maintain 
the single-set structure, the split FWGs obtain empty (zero amplitude) counterparts 
immediately after the split (\fig{fig:spcl} right).  
    
The original implementations of spawning and cloning techniques 
were done for classically moving FWG schemes (\gls{AIMS} and \gls{MCE}) 
and were treating each nuclear FWG separately disregarding that they are part of the total wavefunction.  
The aim of this section is to illustrate rigorous extensions of these ideas to fully quantum TDVP based
formalisms. Here, we will use the diabatic electronic states, 
but extensions to different electronic representations are possible.  

\paragraph{Perturbative spawning (PS):} 
Two main questions that the PS algorithm addresses is {\it when} and {\it where} to spawn? To illustrate 
how these questions are addressed it is instructive to consider two sets of nuclear FWGs representing 
diabatic nuclear wavefunctions (\eq{eq:diBH}) for a two-electronic-state problem (e.g. \fig{fig:spcl} left).
Two conditions need to be satisfied for the PS algorithm to start a spawning sequence: 
1) nuclear components of one or both electronic states reached a nuclear configuration region where 
there is a significant population flow between the electronic states, and 2) the current FWG representation 
does not represent this flow adequately. To assess these two conditions we use perturbative 
estimates of population transfer between the electronic states, where diabatic interstate couplings 
are treated as a perturbation for a zero-order Hamiltonian corresponding to 
diabatic states. The diabatic representation is particularly amenable to the formulation of such estimates 
in the second order of time-dependent perturbation theory because general quadratic forms can be used for 
diabatic potentials and couplings.\cite{Izmaylov:2011ev,Endicott:2014kl}
To obtain a numerical estimate for the first condition 
we evaluate the electronic state population change due to a transition from state $s$ that has 
a nuclear wavefunction $\chi_s(R,t) = \sum_k C_{ks}(t)g_k(R|q_k^{(s)}(t),p_k^{(s)}(t))$ 
to state $s'$ assuming that the receiving state ($s'$) 
can use the {\it complete} basis set of harmonic oscillator eigenstates. This population change 
will be denoted as $P^{(c)}_{s\rightarrow s'}$.
For the second condition, we obtain the same population change but only 
with a restriction that $s'$ can only accept the population from $s$ onto the {\it limited} basis that it has, 
$P^{(l)}_{s\rightarrow s'}$.  Comparing these two estimates, 
$|P^{(l)}_{s\rightarrow s'} - P^{(c)}_{s\rightarrow s'}|>\epsilon$ allows us to make a conclusion whether the 
nuclear basis on $s'$ has enough FWGs to facilitate the population transfer.
If the answer is negative, and spawning is needed, a new FWG is created by optimizing 
its position and momentum to minimize the difference between 
$P^{(l)}_{s\rightarrow s'}$ and $P^{(c)}_{s\rightarrow s'}$.\cite{Izmaylov:2013fe} 

The PS procedure is similar to a perturbative extension of configuration interaction 
space in electronic structure methods such as the B$_{\rm K}$ method.\cite{Bk:1968,Davidson:1981jr} 
The PS approach can also be generalized to locally quadratic diabatization formalisms used in \gls{vMCG}. 
Some form of the diabatic representation is necessary for this spawning technique 
because a local quadratic representation is 
needed for analytical summation over vibrational states to evaluate $P^{(c)}_{s\rightarrow s'}$ quantities. 

\paragraph{Quantum cloning (QC):} To illustrate the QC procedure for the total wavefunction, 
let us consider a case in \fig{fig:spcl} (right) with single-set FWGs on two diabatic electronic states. 
\glspl{EOM} for stacked pairs are obtained using  the action extremum corresponding to \eq{eq:AVP}, 
which results in the following variations for positions and momenta of FWGs encoded in $z_k$'s
\bea
\Re\Bigg{\langle}\delta z_k \frac{\partial\Psi}{\partial z_k}\Bigg{\vert} H-i\partial_t 
\Bigg{\vert}\Psi\Bigg{\rangle} = 0, ~\forall k.
\eea
However, if one allows any pair of stacked FWGs to evolve as components of multi-set,
their evolution will be different because all FWG replicas experience different forces 
on different electronic states. In other 
words, at any moment in time if one considers the total electron nuclear wavefunction with all
CSs pairs within $N_G-1$ single-set and the $k^{\rm th}$ pair to be multi-set 
\bea
\Psi_k(r,R,t) = \sum_s \phi_s(r) \left[C_{ks} g_{k}(R|q_{k}^{(s)}(t),p_{k}^{(s)}(t)) + \sum_{k'\ne k}C_{k's} g_{k'}(R|q_{k'}(t),p_{k'}(t))\right], 
\eea
then because this relaxation (or addition of parameters) came suddenly the variations with respect to 
individual electronic state $z_{k}^{(s)}$ will be non-zero
\bea
\Re\left[ \sum_s \delta z_{k}^{(s)} \Bigg{\langle}\frac{\partial\Psi_k}{\partial z_{k}^{(s)}}
\Bigg{\vert}H-i\partial_t \Bigg{\vert}\Psi_k\Bigg{\rangle}\right]\ne 0.
\eea
This variation contains arbitrary functions $\delta z_{k}^{(s)}(t)$, which can be removed to assess the effect 
of the $k^{\rm th}$ pair split. Naturally, the criterion for splitting the $k^{\rm th}$ pair becomes
the sum over the electronic states
\bea
\sum_s \left| \Re  \Bigg{\langle}\frac{\partial\Psi_k}{\partial z_{k}^{(s)}}
\Bigg{\vert} H-i\partial_t \Bigg{\vert} \Psi_k \Bigg{\rangle}\right| >\epsilon,
\eea
which can be thought as ``decoherence strain''.
This criterion can be applied for any of the $N_G$ single-set pairs at any moment in time. 
It reduces to an intuitive criterion suggested before for an individual pair of CSs and 
accounts for situations when other CS pairs of the full electron-nuclear wavefunction
may reduce the necessity of splitting a particular pair.

Clear advantage of the QC algorithm compare to the PT scheme is use of quantities 
that are already available in regular TDVP algorithms without basis set extensions. This makes QC 
easily applicable not only in the diabatic representation but also in the adiabatic and MCA 
representations. 

\section{Dynamics of open systems}
\label{sec:nosse}

To extend the domain of applicability for TDVP-based FWG methods to even larger systems, 
one can try to generalize these methods to open systems. Here, by an open system we will understand 
a molecular subsystem that is capable of energy but not matter exchange with its environment. 
For example, such systems can be chromophores of photoactive proteins coupled to a 
protein environment (e.g. rhodopsin)~\cite{Hahn:2000/jpcb/1146,Moix:2011/jpcl/3045} 
or a molecule interacting with incoherent light.~\cite{Tscherbul:2014/jpca/3100,Tscherbul:2015/pccp/30904}
Due to a large number of environmental \gls{DOF}, only subsystem dynamics is considered explicitly while 
environmental \gls{DOF} are integrated out.\cite{Nitzan:2006,breuer:2002} This leads to a formalism where
the subsystem density ($\op\rho$) is the main dynamical quantity following the \gls{QME},
$\partial_t\op\rho = \Liou[\op\rho]$ with $\Liou$ as a super-operator modifying the subsystem density.
If the subsystem is uncoupled from the environment, \gls{QME} becomes the Liouville-von Neumann equation, 
$\Liou[\op\rho(t)]=-i[\op H,\op\rho(t)]$, then the subsystem is considered 
to be closed and energy must be conserved.

Combining the McLachlan \gls{TDVP} with \gls{QME} is equivalent to minimizing the error $||\partial_t\op\rho - \Liou[\op\rho]||$ where $||\cdot||$ is the Frobenius (or Hilbert-Schmidt) norm, which results in the following stationary condition~\cite{Raab:2000/tca/358}
\bea\label{eq:TDVP_DM}
\tr\{\delta\op\rho^\dagger(\partial_t\op\rho - \Liou[\op\rho])\} & = & 0.
\eea
Solving \eq{eq:TDVP_DM} using a finite set of moving FWGs leads to two unphysical features 
for the density matrix evolution: 1) the density matrix trace (the subsystem population) is not conserved 
for an open system,~\cite{Gerdts:1997/jcp/3017,Raab:2000/tca/358,Joubert:2015/jcp/134107}
and 2) the energy is not conserved in the closed system limit.~\cite{McLachlan:1964/mp/39,Heller:1976/jcp/63,Raab:2000/tca/358,Joubert:2015/jcp/134107}
The latter leads to an unphysical energy flow channel in the corresponding open system and 
results in wrong dynamics of the subsystem.

Violation of energy conservation for the closed system is not related to non-analyticity of the density 
matrix ansatz since this violation also occurs for analytic parameterizations as well.\cite{Raab:2000/tca/358,Joubert:2015/jcp/134107}
For the sake of simplicity, we will consider an analytic parametrization of the density matrix
(e.g. Bargmann states\cite{Dalton:2014} for the nuclear basis in the diabatic representation). 
The problem of energy flow appear as a consequence of the density matrix entering quadratically 
in \eq{eq:TDVP_DM}, which leads to conservation of the quantity $\tr\{\op\rho^2\op H\}$ for the closed system.
Clearly for mixed states $\rho^2\ne \rho$, and $\tr\{\op\rho^2\op H\}\ne \tr\{\op\rho\op H\}=E$. 
Accounting for this observation, a simple solution to the energy conservation issue would be to propagate 
a square root of the density matrix $\op\rho^{1/2}$ so that the conserved quantity will be energy
\bea
\tr\{\op\rho^2\op H\} & \underset{\op\rho\rightarrow\op\rho^{1/2}}{=} & \tr\{\op\rho\op H\}.
\eea
$\op\rho^{1/2}$ can be seen as an ensemble of states $\{\ket{m_k}\}$ from the density matrix decomposition $\op\rho=\sum_k\ket{m_k}\bra{m_k}$.
An analogue of TDSE for  $\mat m=(\ket{m_1} \ket{m_2} \dots)$ has been given in Ref.~\citenum{Joubert:2014/jcp/234112} and is known as \glsfirst{NOSSE}
\bea\label{eq:NOSSE}
\partial_t\mat m & = & \K[\mat m],
\eea
where $\K$ is obtained as a solution of a $\star$-Sylvester-like equation~\cite{Kressner:2009/na/209}
\bea\label{eq:Sylvester}
\Liou[\op\rho] & = & \mat\K[\mat m]\mat m^\dagger + \mat m\mat\K[\mat m]^\dagger.
\eea
In the limiting case of the closed system, \gls{NOSSE} reduces to a set of uncoupled Schr\"odinger equations.
Therefore, applying the \gls{TDVP} on \gls{NOSSE} is expected to conserve energy as it does for a single Schr\"odinger equation.  
It has been shown that \eq{eq:NOSSE} combined with the McLachlan \gls{TDVP} conserves the energy for the closed system.~\cite{Joubert:2014/jcp/234112}

However, even in \gls{NOSSE}, the subsystem population is still not conserved for a general open system.
To remedy this deficiency, we combined minimization of the \gls{NOSSE} error due to a finite parametrization 
with the Lagrange multipliers method to constrain the subsystem population $\tr\{\mat m\mat m^\dagger\}$, 
the corresponding Lagrangian functional is 
\bea\label{eq:NOSSE_Lag}
\Lambda & = & ||\partial_t\mat m - \mat\K[\mat m]||^2 + \lambda\partial_t\tr\{\mat m\mat m^\dagger\},
\eea
where $\lambda$ is the Lagrange multiplier.
The states $\ket{m_k}$ are expanded in the time-dependent electron-nuclear basis $\{\ket{\varphi_{ks}}\}$
\bea
\ket{m_l} & = & \sum_{k=1}^{N_G}\sum_{s=1}^{N_S} M_{l,ks} \ket{\varphi_{ks}(\mat z_k)},
\eea
where $\{\mat z_k\}$ are parameters for the states $\ket{\varphi_{ks}}$.
In this parameterization, the basis $\{\ket{\varphi_{ks}}\}$ is analytic so that $\partial\ket{\varphi_{ks}}/\partial z_{ka\alpha}^*=0$.
Minimizing \eq{eq:NOSSE_Lag} and solving for the Lagrange multiplier $\lambda$ leads to a set of coupled equations
\bea
\partial_t\mat M & = & \mat S^{-1}[\mat K - \mat\gamma\mat M] -\frac{1}{2}\tr\{\mat K\mat M^\dagger+\mat M\mat K^\dagger\}\mat M, \label{eq:Mdot}\\
\partial_t\mat z & = & \tilde{\mat B}^{-1}\tilde{\mat Y} \label{eq:zdottilde}
\eea
where $\mat S$ and $\mat\gamma$ are matrices defined in \eqs{eq:S} and (\ref{eq:gamma}),
\bea
K_{ks,l} & = & \braket{\varphi_{ks}}{\K_l[\mat m]}, \label{eq:K}\\
\tilde Y_{ka\alpha} & = & \sum_{sl} M_{l,ks}^*\braOket{\frac{\partial\varphi_{ks}}{\partial z_{ka\alpha}}}{\op 1 - \op{\mathcal{P}}}{\K_l[\mat m]}, \label{eq:Ytilde}
\eea
and $\tilde{\mat B}$ is the extension of \eq{eq:B} for an ensemble of analytic states
\bea
\tilde B_{ka\alpha,lb\beta} & = & \sum_{nss'} M_{n,ks}^*\braOket{\frac{\partial\varphi_{ks}}{\partial z_{ka\alpha}}}{\op 1 - \op{\mathcal{P}}}{\frac{\partial\varphi_{ls'}}{\partial z_{lb\beta}}}M_{n,ls'}. \label{eq:Btilde}
\eea

Equations~(\ref{eq:Mdot}) and (\ref{eq:zdottilde}) have been implemented and tested on a simple model system described as a two-state two-dimensional linear vibronic coupling model 
\bea
\op H & = & \sum_{\alpha=1}^2 \frac{\omega_\alpha}{2} \left[R_\alpha^2-\frac{\partial^2}{\partial R_\alpha^2}\right]\sum_{s=1}^{2}\ket{\phi_s}\bra{\phi_s} \nonumber\\
&&+d R_1\Big[\ket{\phi_2}\bra{\phi_2}-\ket{\phi_1}\bra{\phi_1}\Big] \nonumber\\
&&+c R_2\Big[\ket{\phi_2}\bra{\phi_1}+\ket{\phi_1}\bra{\phi_2}\Big],
\eea
where frequencies are $\omega_1=7.743\cdot10^{-3}$ and $\omega_2=6.680\cdot10^{-3}$, and other linear parameters are $d=5.289\cdot10^{-3}$ and $c=9.901\cdot10^{-4}$.
Effect of the external system was accounted for by a non-unitary evolution in the \gls{QME} given in a Lindblad form~\cite{Lindblad:1976/cmp/119,Takagahara:1978/jpsj/728,Wolfseder:1995/cpl/370}
\bea\label{eq:Liouvillian}
\Liou[\op\rho] & = & -i[\op H,\op\rho] + h\sum_{\alpha=1}^2\left[2 \op Z_\alpha\op\rho \op Z_\alpha^\dagger -\op\rho \op Z_\alpha^\dagger\op Z_\alpha -\op Z_\alpha^\dagger\op Z_\alpha\op\rho\right],
\eea
with $h=3.675$, and $\op Z_\alpha$ are defined by
\bea
\op Z_1 & = & \left(R_1-\frac{\partial}{\partial R_1} - \frac{d}{\omega_1}\right)\frac{\ket{\phi_1}\bra{\phi_1}}{\sqrt{2}} + \left(R_1-\frac{\partial}{\partial R_1} + \frac{d}{\omega_1}\right)\frac{\ket{\phi_2}\bra{\phi_2}}{\sqrt{2}}, \label{eq:Z1}\\
\op Z_2 & = & \left(R_2-\frac{\partial}{\partial R_2}\right)\frac{\ket{\phi_1}\bra{\phi_1} + \ket{\phi_2}\bra{\phi_2}}{\sqrt{2}}. \label{eq:Z2}
\eea
This bath is known to violate conservation of density matrix trace along the dynamics,~\cite{Joubert:2015/jcp/134107} and therefore, provides a great test for \gls{NOSSE} combined with the constrained \gls{TDVP}. 
In these simulations the initial density matrix is Boltzmann with temperature $T=1000K$.
The \gls{NOSSE} evolution corresponding to \eq{eq:Liouvillian} is 
\bea
\mat\K[\mat m] & = & -i\op H\mat m + h\sum_{\alpha=1}^2\left[\op Z_\alpha\mat m (\mat m^{-1}\op Z_\alpha\mat m)^\dagger -\op Z_\alpha^\dagger\op Z_\alpha\mat m\right],
\eea
where $\mat m^{-1}=(\mat m^\dagger\mat m)^{-1}\mat m^\dagger$ is a pseudo 
inverse.~\cite{Joubert:2014/jcp/234112}
Calculation of the pseudo inverse can be avoided by reconstructing the density matrix as $\op\rho=\sum_{kl,ss'}\ket{\varphi_{ks}}\rho_{kl,ss'}\bra{\varphi_{ls'}}$ where $\rho_{kl,ss'}=\sum_n M_{n,ks}M_{n,ls'}^*$, and using \eq{eq:Mdot} and \eq{eq:Sylvester} we obtain
\bea\label{eq:rho}
\partial_t\mat\rho & = & \mat S^{-1}\mat L\mat S^{-1} -(\mat S^{-1}\mat\gamma\mat\rho+\mat\rho\mat\gamma\mat S^{-1})-\tr\{\mat S^{-1}\mat L\}\mat\rho, \label{eq:rhodot}
\eea
while \eq{eq:Btilde} and \eq{eq:Ytilde} become
\bea
\tilde B_{k\alpha,l\beta} & = & \sum_{ss'} \braOket{\frac{\partial\varphi_{ks}}{\partial z_{k\alpha}}}{\op 1 - \op{\mathcal{P}}}{\frac{\partial\varphi_{ls'}}{\partial z_{l\beta}}} \rho_{lk,s's}, \label{eq:Btilderho}\\
\tilde Y_{k\alpha} & = & -i\sum_{lss'}\bra{\frac{\partial\varphi_{ks}}{\partial z_{k\alpha}}}(\op 1 - \op{\mathcal{P}})\op H \ket{\varphi_{ls'}}\rho_{lk,s's} \nonumber\\
& & +h\sum_{lss'\beta}\bra{\frac{\partial\varphi_{ks}}{\partial z_{k\alpha}}}(\op 1 - \op{\mathcal{P}})\left[\op Z_\beta\ket{\varphi_{ls'}}[\mat\rho\mat Z_\beta^\dagger\mat S^{-1}]_{lk,s's} - \op Z_\beta^\dagger\op Z_\beta \ket{\varphi_{ls'}}\rho_{lk,s's} \right], \label{eq:Ytilderho}
\eea
where $[\mat Z_\beta]_{kl,ss'}=\braOket{\varphi_{ks}}{\op Z_\beta}{\varphi_{ls'}}$.

\begin{figure}
  \centering
  \includegraphics[width=0.5\textwidth]{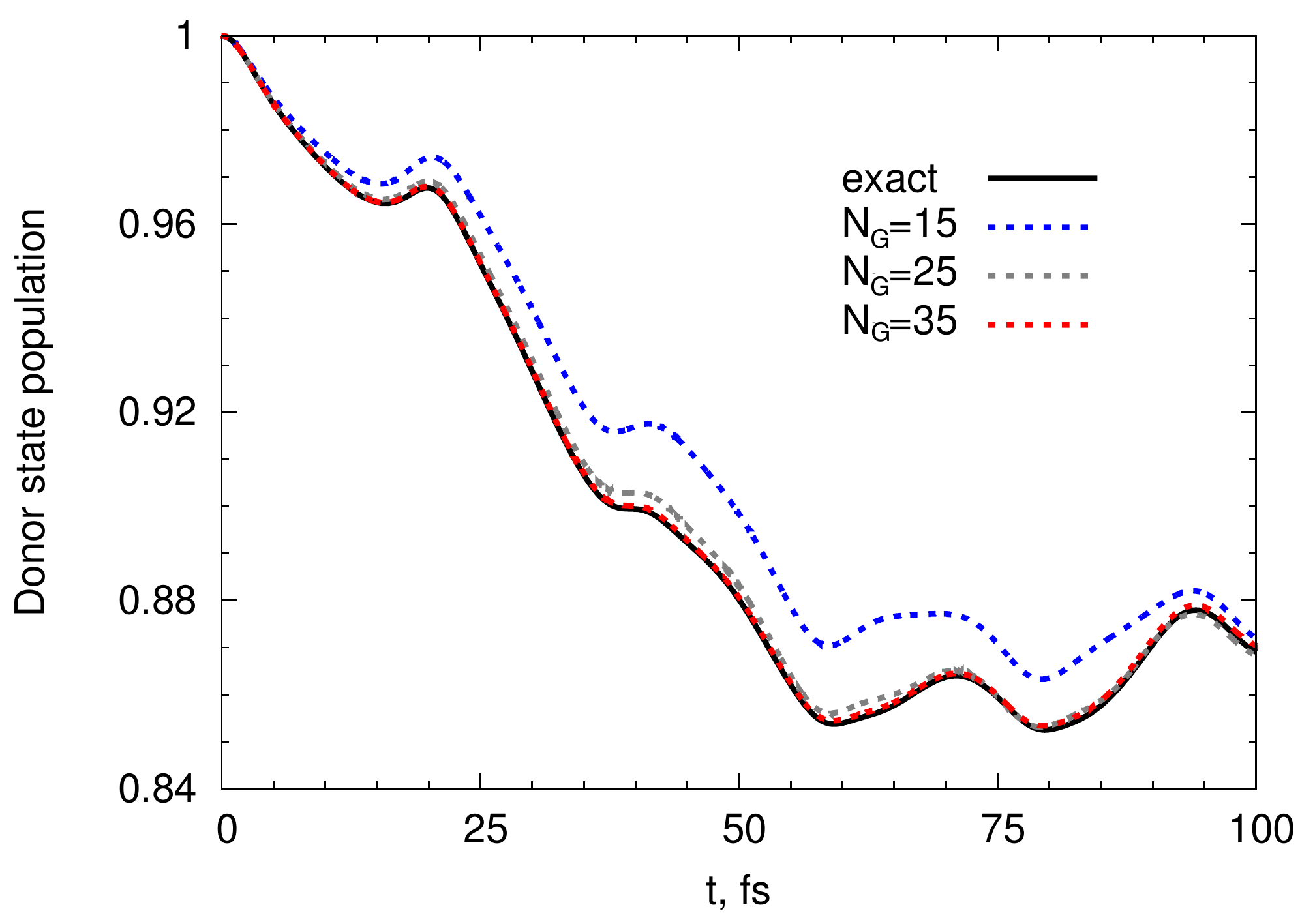}
  \caption{Electronic state population dynamics $\tr\{\op\rho(t)\ket{\phi_1}\bra{\phi_1}\}$ for the exact and constrained \protect\gls{NOSSE} [Eqs.~\eqref{eq:rhodot} and~\eqref{eq:zdottilde}] formalism with different number $N_G$ of basis functions. This figure is reproduced from Ref.~\cite{Joubert:2015/jcp/134107}}
  \label{fig:open}
\end{figure}

Using Eqs.~(\ref{eq:rho}-\ref{eq:Ytilderho}), we demonstrated that combining \gls{NOSSE} with the constrained 
\gls{TDVP} (\eq{eq:NOSSE_Lag}) converges to the exact results with increasing the number 
of nuclear basis functions (\fig{fig:open}). As opposed to the unconstrained approach, 
the population of the subsystem (trace of the density matrix) is conserved in the constraint procedure (Tab.~\ref{tab:open}). In contrast to the \gls{QME} case, the energy is conserved in the closed system limit ($h=0$) 
when \gls{TDVP} is applied to \gls{NOSSE} (Tab.~\ref{tab:close}).

\begin{table}[h!]
  \centering
  \caption{\protect\gls{RMSD} for the trace of the density matrix $\tr\{\op\rho\}$ unsing the constrained \gls{TDVP} [\eq{eq:NOSSE_Lag}] compared to the unconstrained approach for different number $N_G$ of basis functions.}
    \begin{tabular}{@{\hspace{0.5cm}}l@{\hspace{0.5cm}}c@{\hspace{0.5cm}}c@{\hspace{0.5cm}}c@{\hspace{0.5cm}}}
\hline\hline
   $N_G$      &       $15$        &       $25$        &       $35$       \\
\hline
  Constrained & $<10^{-5}$ & $<10^{-5}$ & $<10^{-5}$\\
Unconstrained & $8.8\cdot10^{-2}$ & $2.8\cdot10^{-2}$ & $6.5\cdot10^{-3}$\\
\hline\hline
    \end{tabular}
  \label{tab:open}
\end{table}
\begin{table}[h!]
  \centering
  \caption{\protect\gls{RMSD} of the molecular system energy in units of $\omega_1$ for \gls{TDVP} combined with \protect\gls{NOSSE} [Eqs.~(\ref{eq:rho})-(\ref{eq:Ytilderho})] compared to the usual \protect\gls{QME} approach for different number $N_G$ of basis functions.}
    \begin{tabular}{@{\hspace{0.5cm}}l@{\hspace{0.5cm}}c@{\hspace{0.5cm}}c@{\hspace{0.5cm}}c@{\hspace{0.5cm}}}
\hline\hline
$N_G$      &       $15$        &       $25$        &       $35$       \\
\hline
NOSSE      & $<10^{-5}$ & $<10^{-5}$ & $<10^{-5}$\\
 QME       & $2.8\cdot10^{-3}$ & $2.3\cdot10^{-3}$ & $2.4\cdot10^{-3}$\\
\hline\hline
    \end{tabular}

  \label{tab:close}
\end{table}

\section{Conclusions}
\label{sec:concl}

Wavefunction ansatzes based on linear combinations of moving FWGs guided by the TDVP 
provide powerful techniques for modeling nonadiabatic dynamics of both isolated and 
open molecular systems from first principles. We highlighted several features crucial 
for performance of this set of techniques. First, among possible electronic state representations,
the MCA representation is particularly attractive. It has all the advantages of the 
diabatic representation and can be directly obtained from electronic structure calculations 
as the regular adiabatic representation. The only caveat of the MCA use is
the non-analytic parametrization of the wavefunction. To avoid problems with the system energy 
conservation, this feature requires using the stationary action version of the TDVP. 
Second, if individual nuclear Gaussians are 
propagated classically (or generally independently) the TDVP application to the
FWGs' amplitudes does not conserve the system energy.  
This non-conservation is especially pronounced when several Gaussians interfere in the process 
of dynamics, and it can be related to the basis set incompleteness, which is unavoidable in practice.
The only obvious way to conserve the energy is to propagate parameters of individual 
Gaussians quantum mechanically according to the TDVP in the action formulation. Unfortunately,
this makes equations of motion for Gaussian parameters coupled and increase computational 
complexity of the overall scheme. An interesting question to address is whether there exists 
a scheme for decoupled Gaussian parameters \glspl{EOM} that conserves the system energy? 
Third, to address the increase of the wavefunction complexity in time, a systematic extension of 
the number of nuclear FWGs is needed. Rigorous generalizations of two intuitive schemes 
for nuclear basis set extension (spawning and cloning) has been discussed. 
The spawning technique can be rooted in perturbative estimates 
of population flow between electronic states, while the cloning approach uses estimates of 
the ``decoherence strain'' acting on FWGs moving on different PESs. 
Both schemes can be successfully used for either classically propagated FWGs 
(e.g. \gls{AIMS} and \gls{MCE}) or fully quantum TDVP schemes (e.g. \gls{vMCG}). 
Finally, a straightforward generalization of the TDVP to the Liouville equation in order to 
use FWGs to simulate the reduced density of the open systems encounters 
two fundamental problems: 1) the total energy of the subsystem in a mixed state is not conserved
in the limit of vanishing coupling to the environment, and 2) the subsystem population is not conserved
even in the situations when only the energy exchange is allowed. 
The discussed NOSSE formalism allows to resolve both problems and can be used among 
other applications for first-principle dynamics of molecules under incoherent light conditions. 

\section{Acknowledgments}
This work was supported by Natural Sciences and Engineering Research Council of Canada (NSERC)
through the Discovery Grants Program.

\section{Biographical information}
 
\textbf{Lo{\"{\i}}c Joubert-Doriol} obtained his PhD in 2012 from the University of Montpellier 
under supervision of Fabien Gatti developing models for nonadiabatic quantum dynamics 
of molecules interacting with UV/vis light. As a Marie Curie postdoctoral fellow, he worked 
with Artur Izmaylov and Massimo Olivucci on nonadiabatic 
quantum effects in biological systems. Since 2016, he develops quantum variational approaches 
for nonadiabatic dynamics in Izmaylov's group.

\textbf{Artur Izmaylov} is an Associate Professor at the University of Toronto. 
He received a PhD with Gustavo Scuseria at Rice University in 2008 
and worked as a postdoctoral fellow with John Tully (Yale University) and Michael Frisch (Gaussian Inc.). 
The main efforts of his group are directed toward developing simulation 
techniques for quantum molecular dynamics involving multiple electronic states  
to investigate energy and charge transfer in molecules and nano-structures.

\providecommand*\mcitethebibliography{\thebibliography}
\csname @ifundefined\endcsname{endmcitethebibliography}
  {\let\endmcitethebibliography\endthebibliography}{}

\end{document}